\documentclass[aps,prl,reprint,groupedaddress]{revtex4-1}
\usepackage{graphicx}
\def\vc#1{\mbox{\boldmath $#1$}}
\def\Be{{^{8}{\rm Be}}}
\def\C{{^{12}{\rm C}}}
\def\Ox{{^{16}{\rm O}}}
\def\Ne{{^{20}{\rm Ne}}}

\begin{document}


\title{Container evolution for cluster structures in $\Ox$}


\author{Y.~\textsc{Funaki}}
\affiliation{
College of Science and Engineering, Kanto Gakuin University, Yokohama 236-8501, Japan \\
School of Physics and Nuclear Energy Engineering and IRCNPC, Beihang University, Beijing 100191, China \\
Nishina Center for Accelerator-Based Science, RIKEN, Wako 351-0198, Japan}


\date{\today}

\begin{abstract}
\begin{description}
\item[Background]
$\alpha+\C$ clustering in $\Ox$ has been of historical importance in nuclear clustering. In the last 15 years the $4\alpha$ condensate state has been proposed as a new-type cluster state.
\item[Purpose]
The aim is to reveal a dynamical process of the formation of different kinds of cluster states, in terms of a ``container'' aspect of clusters, in $\Ox$.
\item[Method]
The so-called THSR wave function for the $4\alpha$ clusters is extended to inclusion of two different containers occupied independently by the $\C$ $(3\alpha)$ and $\alpha$ clusters.
\item[Results]
The five $J^\pi=0^+$ states with $4\alpha$ tetrahedral shape, $\alpha+\C$ cluster structures, and the $4\alpha$ condensate character, are found to be represented, to good approximation, by single configurations of the extended THSR wave function with containers of appropriate shape and size.
\item[Conclusions]
It is demonstrated in $\Ox$ that the dynamical evolution of cluster structures can be caused by size and shape evolution of a container occupied with clusters. The $\alpha$ condensate with gaslike $4\alpha$ configuration appears as a limit of the cluster formation.
\end{description}
\end{abstract}


\maketitle

Alpha-like four-nucleon correlation plays an important role in nuclei, in which spin and isospin are saturated. In particular, $N=Z$ light nuclei tend to have $\alpha$ cluster structures in their excited states. The $3\alpha$ cluster structure in $\C$, $\alpha+\C$ cluster structure in $\Ox$ and $\alpha+\Ox$ cluster structure in $\Ne$ are typical examples and their realities are firmly established in many historical works~\cite{Ik80}.

In the past 15 years, alpha-particle condensate structure has been extensively studied theoretically and experimentally. Although providing direct observatory evidence is still in an open question~\cite{Ra13,Fr14,It14}, many theoretical calculations predict the existence of the $3\alpha$ and $4\alpha$ condensate states in $\C$ and $\Ox$, respectively, in which all $\alpha$ clusters weakly interact with each other with a dilute gaslike configuration, and occupy an identical orbit of a meanfield-like potential~\cite{To01,Fu03,Ya04,Ya05,Fu08}. 

On the other hand, the ordinary non-gaslike cluster states like the $\alpha+\Ox$, $\alpha+\C$ inversion doublets, linear-chain $\alpha$-cluster states, etc. are completely different from the gaslike cluster states. They had been understood by a concept of localized clustering, in which all clusters are in a geometric arrangement.  However, more recent works have required us to modify the basic idea of understanding the ordinary cluster states. The authors in Ref.~\cite{Zh13} introduced a microscopic $\alpha+\Ox$ cluster model wave function, which demonstrates a nonlocalized motion of the $\alpha$ and $\Ox$ clusters. They proved that the model wave function coincides with the full solution of $\alpha+\Ox$ RGM equation of motion for all the $\alpha+\Ox$ inversion doublet band states. Similar results are also obtained for $3\alpha$ and $4\alpha$ linear-chain states~\cite{Su14}, which are originally proposed by Morinaga~\cite{Mo56}. All these results lead to the idea that dynamically mutual clusters are confined in a ``container'', whose shape and size are flexibly conformed, in a nonlocalized way.
This new concept of the so-called ``container'' picture modifies the preceding understanding of nuclear clustering, since the localized clustering has been an important basis to understand the ordinary (non-gaslike) nuclear cluster structures. The spatial localization of clusters seems to appear when the size of container is very small, due to the effect of Pauli principle acting on clusters in between, as a kinematical effect. 


In this Letter, I discuss the nuclear clustering in $\Ox$, since this is the typical nucleus of gaslike and non-gaslike cluster states coexisting. The special interest is in how both gaslike and non-gaslike cluster states, which seem to be quite different from each other, are successively formed as the increase of excitation energy.
The $\alpha+\C$ cluster structure in $\Ox$ is formed by the activation of cluster degree of freedom in the ground state having a dual property~\cite{Ya08,Fr17,Zh17}, i.e. by the excitation of relative motion between the $\alpha$ and $\C$ clusters. The gaslike $4\alpha$ cluster state is then generated as a result of further excitation of the $\C$ core, to the $3\alpha$ cluster state, i.e. to the Hoyle state~\cite{Fr14}. The path of this cluster evolution is shown in the famous Ikeda diagram, together with many other paths in many other nuclei~\cite{Ik68}. I show that this path of cluster evolution along the excitation energy in $\Ox$ is nothing but the path of a size and shape evolution of a ``container'', which provides a new framework of describing both gaslike and non-gaslike cluster states simultaneously.

I adopt a model wave function to realize the above mentioned picture, which is the extended version of the so-called THSR wave function (eTHSR). It is given by a natural extension of what are used in many previous works of $\Be$~\cite{Fu02}, $\C$~\cite{To01,Fu03,Fu05,Zh14,Fu15b,Zh16}, $\Ox$~\cite{Fu10}, $\Ne$~\cite{Zh13}, ${^9{\rm Be}}$~\cite{Ly15} and ${^{10}{\rm Be}}$~\cite{Ly16}, ${^9_\Lambda {\rm Be}}$~\cite{Fu14}, ${^{13}_\Lambda {\rm C}}$~\cite{Fu17}, etc, and has the following form:
\begin{eqnarray}
&&\Phi(\vc{\beta}_1, \vc{\beta}_2) \nonumber \\
&& = {\cal A}\Big[ \exp \Big\{ -\hspace{-0.1cm} \sum_{k}^{x,y,z}\hspace{-0.1cm} \frac{1}{2B_{1k}^2} \Big(\mu_1\xi_{1k}^2+\mu_2\xi_{2k}^2 \Big) \Big\} \phi(\alpha_1)\phi(\alpha_2)\phi(\alpha_3) \nonumber \\
&& \times \exp \Big\{ -\hspace{-0.1cm} \sum_{k}^{x,y,z}\hspace{-0.1cm} \frac{1}{2B_{2k}^2}\mu_3\xi_{3k}^2 \Big\}  \phi(\alpha_4) \Big],\nonumber \\
\label{eq:thsr} 
\end{eqnarray}
with ${\cal A}$ being the antisymmetrization operator acting on the 16 nucleons, $\phi(\alpha_i)$ the internal wave function of the $i$-th $\alpha$ particle assuming a $(0s)^4$ configuration, like,
\begin{equation}
\phi(\alpha_i)\propto \exp\big[-\sum_{1\leq j<k \leq 4}(\vc{r}_{4(i-1)+j}-\vc{r}_{4(i-1)+k})^2/(8b^2)\big],
\end{equation}
$\vc{\xi}_i=\vc{R}_{i+1}-(\vc{R}_1+\cdots+\vc{R}_i)/i$, the Jacobi coordinates between the $\alpha$ particles with $\vc{R}_i=\sum_{j=1}^{4}\vc{r}_{4(i-1)+j}/4$ the position vectors of the $i$-th $\alpha$ particle, and $\mu_i=4i/(i+1)$, for $i=1,2,3$. The parameter $b$ characterizes the size of the constituent $\alpha$ particle, while the parameters $\vc{B}_1$ and $\vc{B}_2$ characterize the size and shape of a container, in which the $\alpha$ clusters are confined. I can instead define parameters $\vc{\beta}_i$ that satisfies the relation, $B_{jk}^2=b^2+2\beta_{jk}^2$, with $j=1,2$ and $k=x,y,z$. However, throughout this study, I assume the axial symmetry $\beta_{i\perp}\equiv \beta_{ix}=\beta_{iy}$, so as to deal with the four parameters, $\beta_{1\perp}, \beta_{1z}, \beta_{2\perp}, \beta_{2z}$, in the practical calculations.

The exponential functions in Eq.~(\ref{eq:thsr}) represent the center-of-mass (c.o.m.) motions of the $\alpha$ clusters, in terms of the corresponding Jacobi coordinates. If the $\vc{B}_1$ and $\vc{B}_2$ take a common value, i.e. $\vc{B}_1=\vc{B}_2=\vc{B}$, then Eq.~(\ref{eq:thsr}) results in,
\begin{equation}
\Phi(\vc{\beta})={\cal A} \Big[ \prod_{i=1}^4 \exp \Big\{- 2\sum_{k}^{x,y,z} (R_{ik}-X_{Gk})^2/B_{k}^2 \Big\}\phi(\alpha_i) \Big],
\end{equation}
with $\vc{X}_G$ the total c.o.m. coordinate, the original THSR wave function, in which all $\alpha$ clusters occupy an identical orbit. This is the $\alpha$ condensate state in a ``gas'' phase if the magnitude of the parameter $|\vc{B}|$ is large enough for the antisymmetrizer ${\cal A}$ to be negligible~\cite{Fu09}.

\begin{figure}[htbp]
\begin{center}
\includegraphics[scale=0.65, bb=-75 620 80 797]{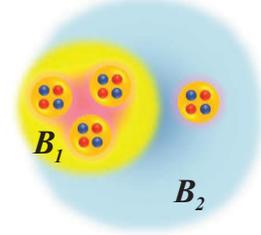}
\caption{(Color online) Schematic representation of the eTHSR wave function, in which the two containers of the $3\alpha$ and $\alpha$ clusters are characterized by the parameters $\vc{B}_1$ and $\vc{B}_2$, respectively.}
\label{fig:ethsr}
\end{center}
\end{figure}

We should note that the way of describing cluster states in this model wave function is completely different from those in other traditional cluster models, like the Brink-Bloch wave function~\cite{Br66} and even the Antisymmetrized Molecular Dynamics (AMD) wave function~\cite{Ka95}, in which clusters are spatially positioned in a localized way, to form a multi-centered Slater determinant. In the present model, constituent clusters are arranged without mutually forming any geometric rigid-shaped configuration. This wave function is then far away from the single multi-centered Slater determinant and is represented as infinite number of superposition of the Slater determinants~\cite{Fu15a}. Very schematic picture representing the eTHSR wave function in the present $\Ox$ system is shown in FIG.~\ref{fig:ethsr}, in which the $3\alpha$ clusters and another $\alpha$ cluster are confined in different containers characterized by the parameters $\vc{B}_1$ and $\vc{B}_2$, respectively. This is contrasted with the Brink-Bloch wave function, in which the cluster configurations are described by their relative distance parameters.

In order to obtain the energy spectrum, first I impose the so-called $r^2$-constraint method~\cite{Fu06,Fu10,Fu15b,Zh16}, to effectively and roughly eliminate spurious continuum components from the present model space. A spurious continuum state is calculated to have a large r.m.s. radius in the bound state approximation, and hence this method is to remove the components with extremely large r.m.s. radii in the following way, 
\begin{eqnarray}
&&\hspace{-0.4cm} \sum_{\vc{\beta}_1^\prime,\vc{\beta}_2^\prime} \langle \Phi^{J=0}(\vc{\beta}_1,\vc{\beta}_2) |{\widehat {\cal O}}_{\rm rms}- \{R^{(\gamma)}\}^2 |\Phi^{J=0}(\vc{\beta}_1^\prime,\vc{\beta}_2^\prime) \rangle \nonumber \\ 
&& \times g^{(\gamma)}(\vc{\beta}_1^\prime,\vc{\beta}_2^\prime)=0, \label{eq:cutoff1}
\end{eqnarray}
with ${\widehat {\cal O}}_{\rm rms} = \sum_{i=1}^{16}(\vc{r}_i-\vc{X}_G)^2/16$, and $\Phi^{J=0}(\vc{\beta}_1,\vc{\beta}_2)={\widehat P}^{J=0} \Phi(\vc{\beta}_1,\vc{\beta}_2)$, where ${\widehat P}^{J=0}$ is the projection operator of angular-momentum $J=0$. The eigenfunctions of the above equation are expressed below,
\begin{equation}
\Phi^{(\gamma)} = \sum_{\vc{\beta}_1,\vc{\beta}_2}g^{(\gamma)}(\vc{\beta}_1,\vc{\beta}_2) \Phi^{J=0}(\vc{\beta}_1,\vc{\beta}_2). \label{eq:cutoff2}
\end{equation}
I now eliminate the eigenstates with the eigenvalues, $R^{(\gamma)} \ge 7.0\ {\rm fm}$, from the following linear combination:
\begin{equation}
\Psi_{\lambda} = \sum_{\gamma} f_{\lambda}^{(\gamma)} \Phi^{(\gamma)}. \label{eq:wf}
\end{equation}
The coefficients of the above expansion is determined by solving the Hill-Wheeler equation,
\begin{equation}
\sum_{\gamma^\prime} \langle \Phi^{(\gamma)} |H-E_\lambda| \Phi^{(\gamma^\prime)} \rangle f_{\lambda}^{(\gamma^\prime)} = 0. \label{eq:eigenwf}
\end{equation}

For Hamiltonian, I adopt the effective nucleon-nucleon interaction with finite range three-body force called F1 force~\cite{To94}. Note that for example if we adopt other forces like Volkov No.1 and No.2~\cite{Vo65}, no reasonable force parameter is found to adjust $\alpha+\C$ threshold energy and the ground state energy of $\Ox$~\cite{It95}, while it is shown that the F1 force gives much better description of $\C$ and $\Ox$~\cite{It16}.

\begin{figure}[htbp]
\begin{center}
\includegraphics[scale=1.0]{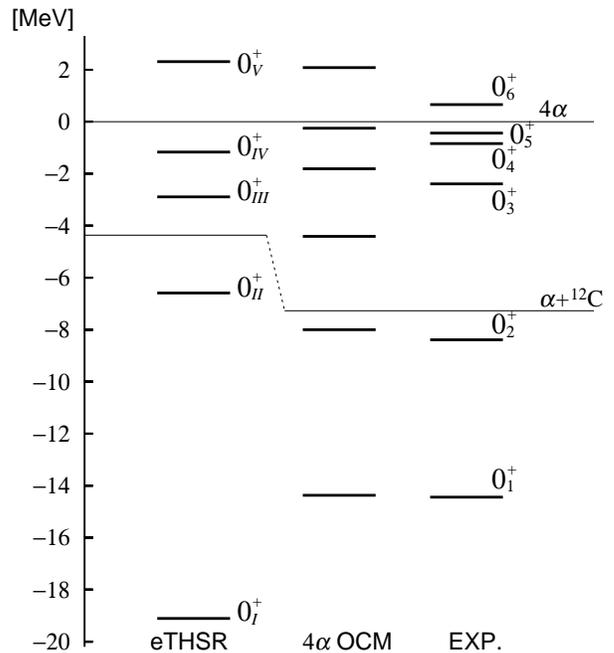}
\caption{Energy spectra of the low-lying $J^\pi=0^+$ states calculated with the extended THSR ansatzes. The corresponding observed spectrum (Exp.)~\cite{Aj86} and result by the $4\alpha$ OCM~\cite{Fu08} are also shown. The $0_4^+$ state in experiment is taken from Ref.~\cite{Wa07}.}
\label{fig:level}
\end{center}
\end{figure}

In FIG.~\ref{fig:level}, the calculated energy spectrum for $J^\pi=0^+$ states is shown. The corresponding experimental data and result by the previous $4\alpha$ OCM calculation~\cite{Fu08} are also shown. The solution of Hill-Wheeler equation with the $r^2$ constraint method is shown. The $0_{V}^+$ state is actually the seventh $0^+$ state obtained by solving the Hill-Wheeler equation, i.e. the fifth and sixth eigenstates are kicked out from the present consideration, since they have larger r.m.s. radii and are regarded as spurious continuum states accidentally mixed with the physical states. 

In the $4\alpha$ OCM calculation, it is reported that the $0_6^+$ state has the $4\alpha$ condensate character and the $0_2^+$ - $0_5^+$ states all have $\alpha+\C$ cluster structures. i.e. $\alpha(S)+\C(0_1^+)$, $\alpha(D)+\C(2_1^+)$, $\alpha(S)+\C(0_1^+)$, and $\alpha(P)+\C(1^-)$ cluster structures, respectively. The difference between the $0_2^+$ and $0_4^+$ states are that in the latter the $\alpha$ and $\C$ relative motion is excited and has a higher nodal $S$-wave, to have a larger r.m.s. radius than the former. 

Since in the present eTHSR wave function of Eq.~(\ref{eq:thsr}) the $\alpha$ clusters occupy positive parity orbits, such a state as having the $\alpha(P)+\C(1^-)$ cluster structure, like the $0_5^+$ state in the OCM calculation, is missing. The inclusion of negative parity orbit in the THSR ansatz is also possible and will be shown in the forthcoming paper. I mention that in fact an extension to such a direction is already done~\cite{Zh13,Ly15,Ly16}. However, for the other states, a one-to-one correspondence to the experimental data as well as to the $4\alpha$ OCM calculation is consistently obtained. 

\begin{table}[htbp]
\begin{center}
\caption{R.m.s. charge radii and monopole matrix elements of the $0_{I}^+$ - $0_{V}^+$ states calculated with the eTHSR ansatz, in comparison with the corresponding experimental data.}\label{tab:rms}
\begin{tabular}{ccccc}
\hline\hline
 & \multicolumn{2}{c}{eTHSR} & \multicolumn{2}{c}{Exp.} \\
 & $R_{\rm rms}$ (fm) & $M(E0)$ $(e {\rm fm}^2)$ & $R_{\rm rms}$ (fm) & $M(E0)$ $(e {\rm fm}^2)$ \\
\hline
$0_I^+$ & $2.71(0.02)$ &  & $2.70$ &  \\
$0_{II}^+$ & $3.2$ & $5.9$ &  & $3.55(0.21)$ \\
$0_{III}^+$ & $3.3$ & $5.7$ &  & $4.03(0.09)$ \\
$0_{IV}^+$ & $4.9$ & $0.8$ &  &  \\
$0_{V}^+$ & $4.9$ & $0.7$ &  &  \\
\hline\hline
\end{tabular}
\end{center}
\end{table}
In TABLE~\ref{tab:rms}, r.m.s. radii and monopole matrix elements with the ground state are shown. The experimental data available are reasonably reproduced. We can also see that from the $0_I^+$ to the $0_{V}^+$ states, i.e. as the states are excited, the r.m.s. radius becomes larger and the monopole matrix element becomes smaller. This indicates that the higher the excitation energy is, the more evolved the clustering is. The evolution of the clustering can be described by solving the Hill-Wheeler equation concerning the model parameters $\vc{\beta}_1$ and $\vc{\beta}_2$.

\begin{figure*}[htbp]
\begin{center}
\includegraphics[scale=0.85]{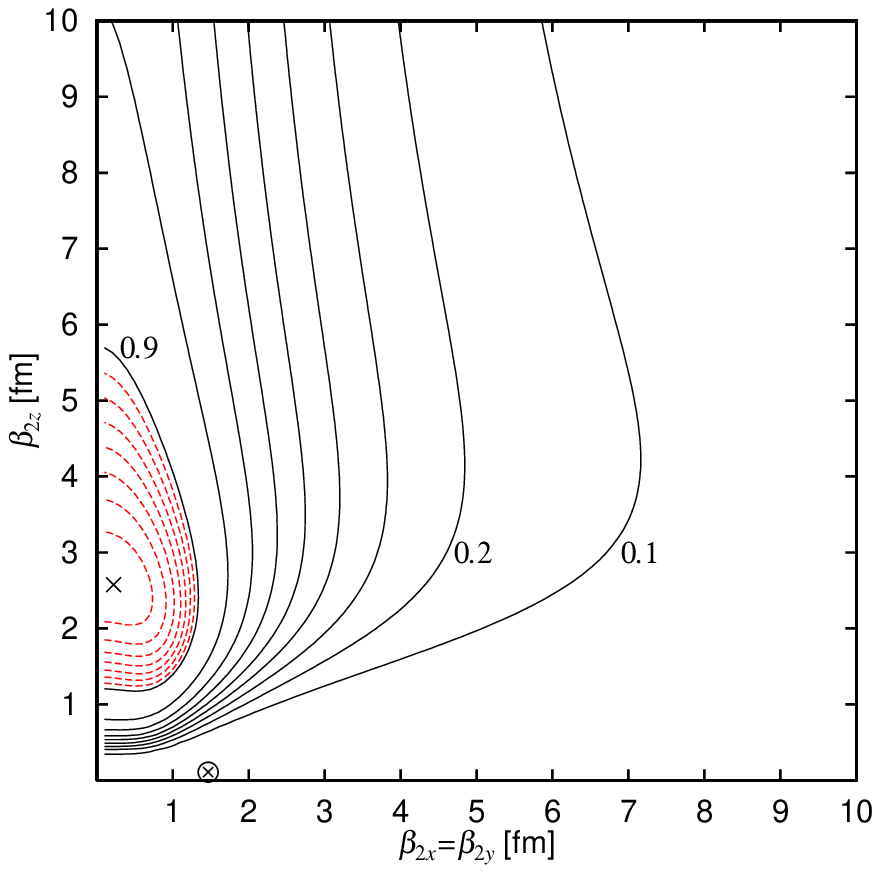}
\includegraphics[scale=0.85]{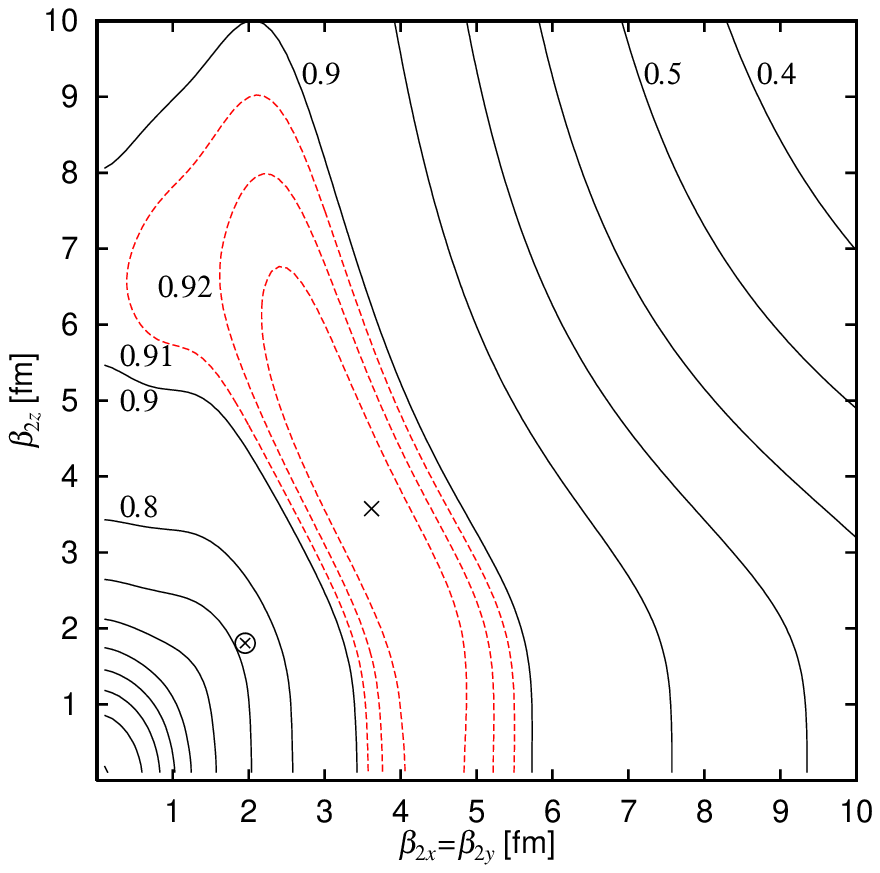}
\includegraphics[scale=0.85]{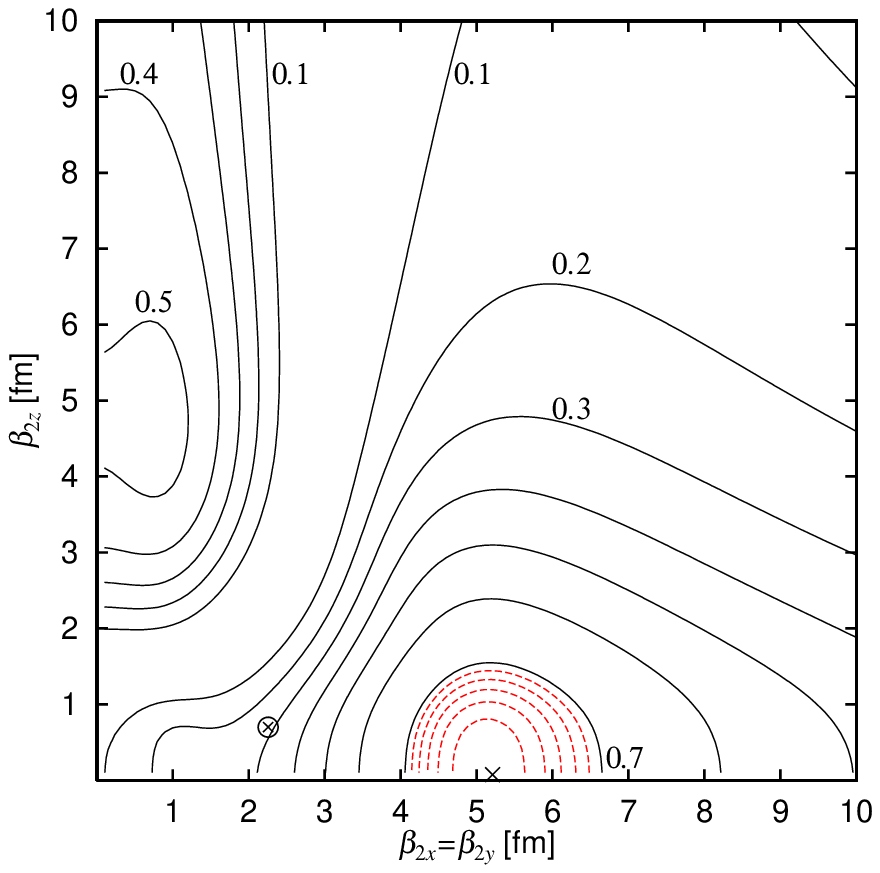}
\includegraphics[scale=0.85]{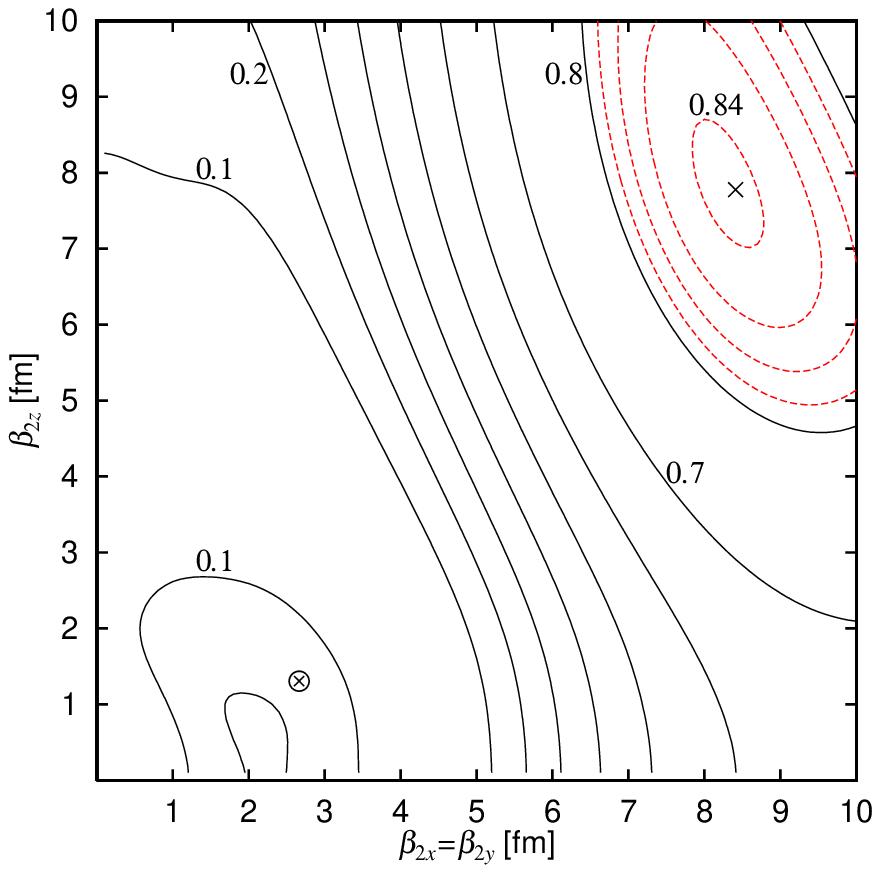}
\caption{(Color online) Contour maps of the squared overlaps between the $0_{I}^+$ (left top), $0_{II}^+$ (right top), $0_{III}^+$ (left bottom), and $0_{IV}^+$ (right bottom) states, and the single extended deformed THSR wave functions, in two-parameter space $\beta_{2x}=\beta_{2y}$ and $\beta_{2z}$, in which $\vc{\beta}_1$ parameter values are fixed at optimal ones, denoted by $\otimes$, so that the maxima in four-parameter space $\beta_{1x}=\beta_{1y},\ \beta_{1z},\ \beta_{2x}=\beta_{2y},\ \beta_{2z}$ appear in these figures. The maximum positions are denoted by $\times$. Red dotted contour lines are in a step of $0.01$ and Black solid ones are in a step of $0.1$.}
\label{fig:so1-4}
\end{center}
\end{figure*}

\begin{figure}[htbp]
\begin{center}
\includegraphics[scale=0.85]{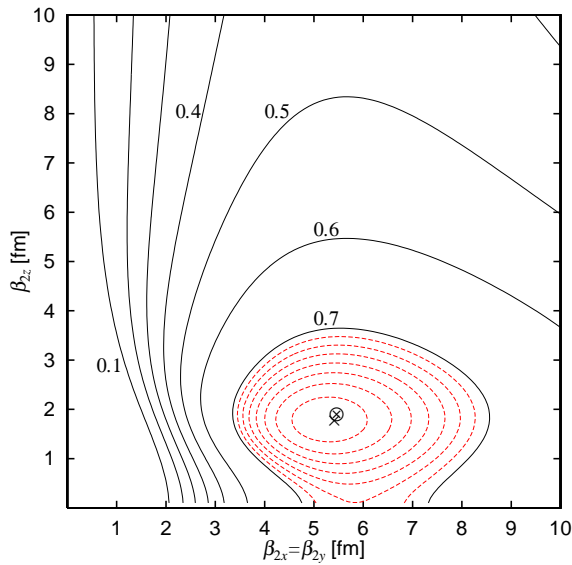}
\caption{(Color online) Contour map of the squared overlap for the $0_{V}^+$ state, shown in the same way as in FIG.~\ref{fig:so1-4}.}
\label{fig:so5}
\end{center}
\end{figure}

This respect is made much clearer by calculating the following squared overlap: 
\begin{equation}
{\cal O}_{\lambda}(\vc{\beta}_1,\vc{\beta}_2)=|\langle {\widetilde \Phi}^{J=0}_{\lambda}(\vc{\beta}_1,\vc{\beta}_2) |\Psi_{\lambda} \rangle|^2, \label{eq:so}
\end{equation}
where ${\widetilde \Phi}^{J=0}_{\lambda}(\vc{\beta}_1,\vc{\beta}_2)$ is normalized single eTHSR wave function in a space orthogonal to the lower eigenstates, like ${\widetilde \Phi}^{J=0}_{I}(\vc{\beta}_1,\vc{\beta}_2)={\cal N}_{I}\Phi^{J=0}(\vc{\beta}_1,\vc{\beta}_2)$, ${\widetilde \Phi}^{J=0}_{II}(\vc{\beta}_1,\vc{\beta}_2)={\cal N}_{II}(1-|\Psi_{I} \rangle\langle \Psi_{I}|)\Phi^{J=0}(\vc{\beta}_1,\vc{\beta}_2)$, ${\widetilde \Phi}^{J=0}_{III}(\vc{\beta}_1,\vc{\beta}_2)={\cal N}_{III}(1-|\Psi_{I} \rangle\langle \Psi_{I}|-|\Psi_{II} \rangle\langle \Psi_{II}|)\Phi^{J=0}(\vc{\beta}_1,\vc{\beta}_2)$, etc., with ${\cal N}_{I}$, ${\cal N}_{II}$, $\cdots$, the corresponding normalization constants.

This quantity indicates how these five states $\Psi_{\lambda}$ $(\lambda=I,\cdots,V)$ are expressed by single configurations of the eTHSR wave functions, and therefore, gives direct information of whether the container structure is realized or not in these states, and if so, what kind of containers represent the states.
In FIGs.~\ref{fig:so1-4} and \ref{fig:so5}, the contour maps of squared overlap of the states $\Psi_{I}$-$\Psi_{V}$ with single configurations in the $\beta_{2\perp}$ and $\beta_{2z}$ parameter space in Eq.~(\ref{eq:so}) are shown. Here the $\vc{\beta}_1$ parameter, i.e. $(\beta_{1\perp},\beta_{1z})$ is fixed at the position denoted by $\otimes$ in these figures, so that the maximum value of the squared overlap in the four parameter space $(\beta_{1\perp},\beta_{1z},\beta_{2\perp},\beta_{2z})$ appears at the position denoted by $\times$. The maximal values and $\vc{\beta}_1$ and $\vc{\beta}_2$ parameter values to give the maxima are listed in TABLE~\ref{tab:somax}. The corresponding $\vc{B}_1$ and $\vc{B}_2$ values are also shown.

Before discussing the features of the $0_{I}^+$ - $0_{V}^+$ states, I show in TABLE~\ref{tab:somax_c} the maximum values of the squared overlap of the $0_1^+$, $2_1^+$ and $0_2^+$ states in $\C$ with the single $3\alpha$ THSR configuration. This is the same calculations as those for $\Ox$. The states in $\C$ are calculated with the $3\alpha$ THSR ansatz with the same F1 force parameters. It is now well known that these states are very precisely described by single THSR configurations, compared with the $3\alpha$ RGM and $3\alpha$ GCM~\cite{Fu80}, not only for the Hoyle state $(\C(0_2^+))$ but also for the other $0_1^+$ and $2_1^+$ states~\cite{Fu05}. All these very large squared overlap values shown in this table again mean that the present container picture nicely holds not only for the dilute gaslike Hoyle state but also for the much more compact $0_1^+$ and $2_1^+$ states. 

\begin{table}[htbp]
\begin{center}
\caption{Maxima of the squared overlaps in FIGs.~\ref{fig:so1-4},~\ref{fig:so5} for the $0_{I}^+$ - $0_{V}^+$ states, in the four-parameter space $(\beta_{1\perp},\beta_{1z},\beta_{2\perp},\beta_{2z})$. The corresponding parameter values $(B_{1\perp}, B_{1z},B_{2\perp}, B_{2z})$ are also shown.}\label{tab:somax}
\begin{tabular}{ccclcl}
\hline\hline
 & ${\cal O}_{max}$ & \multicolumn{2}{c}{$(\beta_{1\perp},\ \beta_{1z},\ \beta_{2\perp}\,\beta_{2z})$} & \multicolumn{2}{c}{$(B_{1\perp},\ B_{1z},\ B_{2\perp},\ B_{2z})$} \\
\hline
$0_{I}^+$ & $0.98$ & \multicolumn{2}{c}{$(1.3,\ 0.1,\ 0.1,\ 2.6\ {\rm fm})$} & \multicolumn{2}{c}{$(2.3,\ 1.4,\ 1.4,\ 3.9\ {\rm fm})$} \\
$0_{II}^+$ & $0.94$ & \multicolumn{2}{c}{$(1.8,\ 1.8,\ 3.5,\ 3.6\ {\rm fm})$} & \multicolumn{2}{c}{$(2.9,\ 2.9,\ 5.2,\ 5.3\ {\rm fm})$} \\
$0_{III}^+$ & $0.76$ & \multicolumn{2}{c}{$(2.1,\ 0.7,\ 5.1,\ 0.1\ {\rm fm})$} & \multicolumn{2}{c}{$(3.3,\ 1.7,\ 7.4,\ 1.4\ {\rm fm})$} \\
$0_{IV}^+$ & $0.84$ & \multicolumn{2}{c}{$(2.5,\ 1.3,\ 8.3,\ 7.8\ {\rm fm})$} & \multicolumn{2}{c}{$(3.8,\ 2.3,\ 11.8,\ 11.1\ {\rm fm})$} \\
$0_{IV}^+$ & $0.78$ & \multicolumn{2}{c}{$(5.3,\ 1.9,\ 5.3,\ 1.8\ {\rm fm})$} & \multicolumn{2}{c}{$(7.6,\ 3.0,\ 7.6,\ 2.9\ {\rm fm})$} \\
\hline\hline
\end{tabular}
\end{center}
\end{table}

\begin{table}[htbp]
\begin{center}
\caption{Maxima of the squared overlaps for the $0_1^+$, $2_1^+$ and $0_2^+$ states in $\C$ in two-parameter space $\beta_{\perp}$ and $\beta_{z}$. The corresponding $B_{\perp}$ and $B_{z}$ values are also shown.}\label{tab:somax_c}
\begin{tabular}{ccclcl}
\hline\hline
 & ${\cal O}_{max}$ & \multicolumn{2}{c}{$(\beta_{\perp},\ \beta_{z})$} & \multicolumn{2}{c}{$(B_{\perp},\ B_{z})$} \\
\hline
$\C(0_1^+)$ & $0.93$ & \multicolumn{2}{c}{$(1.9,\ 1.8\ {\rm fm})$} & \multicolumn{2}{c}{$(3.0,\ 2.9\ {\rm fm})$} \\
$\C(2_1^+)$ & $0.90$ & \multicolumn{2}{c}{$(1.9,\ 0.5\ {\rm fm})$} & \multicolumn{2}{c}{$(3.0,\ 1.6\ {\rm fm})$} \\
$\C(0_2^+)$ & $0.99$ & \multicolumn{2}{c}{$(5.6,\ 1.4\ {\rm fm})$} & \multicolumn{2}{c}{$(8.0,\ 2.4\ {\rm fm})$} \\
\hline\hline
\end{tabular}
\end{center}
\end{table}

Then, let us investigate the features for all these $0_I^+$ - $0_{V}^+$ states one by one.

In the ground state, shown in FIG.~\ref{fig:so1-4}(left top), $3\alpha$ clusters are put into an oblately deformed and very compact container with $\beta_{1\perp}\gg \beta_{1z}$, while the remaining $\alpha$ cluster is put into a prolately deformed and very compact container with $\beta_{2\perp}\ll\beta_{2z}$. This means that the first $3\alpha$ clusters move in a $xy$-plane and the last $\alpha$ cluster moves in $z$-direction. This supports the idea that the ground state has a tetrahedral shape of the $4\alpha$ clusters proposed by several authors~\cite{Bi14,Ka17}. Our calculation indicates that this configuration is contained in the $0_I^+$ state by $98$ \%.

In the $0_{II}^+$ state, shown in FIG.~\ref{fig:so1-4}(right top), the $3\alpha$ clusters are in a spherical container with $\beta_{1\perp}\sim \beta_{1z}$. The fourth $\alpha$ cluster is put into a larger size container with spherical shape, i.e. $\beta_{2\perp}\sim \beta_{2z} > \beta_{1\perp}\sim \beta_{1z}$. In particular, the parameter set $(\beta_{1\perp},\beta_{1z})=(1.8,1.8\ {\rm fm})$ is almost the same as that for $\C$ in TABLE~\ref{tab:somax_c}, i.e. $(\beta_{\perp},\beta_{z})=(1,9,1.8\ {\rm fm})$. This means that the first $3\alpha$ clusters are confined in a compact container to form the ground state of $\C$, since the $\C(0_1^+)$ state can be very precisely described by the single configuration with these parameter values. The fourth $\alpha$ cluster moves in a larger spherical container, because of $(\beta_{2\perp},\beta_{2z})=(3.5,3.6\ {\rm fm})$, which gives the largest squared overlap $94$ \%. This is the new interpretation of the $\alpha+\C$ cluster structure, whose traditional understanding is that the $\alpha$ cluster orbits in an $S$-wave around the $\C(0_1^+)$ state. 

The $0_{III}^+$ state, which is shown in FIG.~\ref{fig:so1-4}(left bottom), is similar to the $0_{II}^+$ state but both containers are not spherical but deformed. The $\vc{\beta}_1$ parameter takes almost the same value as that of the isolated $\C(2^+)$ state, as shown in TABLE~\ref{tab:somax_c}, which means that the first $3\alpha$ clusters form the $\C(2^+)$ state, since the state is described by the single parameter value of $\vc{\beta}$. The configuration of the remaining $\alpha$ cluster $(\beta_{2\perp},\beta_{2z})=(5.1,0.1\ {\rm fm})$, giving the largest value $76$ \%, means that the $\alpha$ cluster moves in a deformed and larger container. This is present understanding of the $0_3^+$ state, which is conventionally considered to have the $\alpha(D)+\C(2^+)$ structure. 

In the $0_{IV}^+$ state, shown in FIG.~\ref{fig:so1-4}(right bottom), one can see that the $3\alpha$ clusters are put in slightly larger container than that for the $\C(0_1^+)$ state, which is slightly deformed in a oblate shape. The fourth $\alpha$ cluster, however, moves in a much larger and almost spherical container, like a satellite. This configuration expresses the $0_{IV}^+$ state dominantly by $84$ \%. This means that the second container characterized by $\vc{\beta}_2$ is further evolved from that in the $0_{II}^+$ state. I can say that this state corresponds to the $0_4^+$ state in the former $4\alpha$ OCM calculation, which predicts the $\alpha+\C(0_1^+)$ higher nodal structure for the state.

The $0_{V}^+$ state, shown in FIG.~\ref{fig:so5}, is the most striking. All the $\alpha$ clusters occupy an identical orbit, with $(\beta_{1\perp},\beta_{1z},\beta_{2\perp},\beta_{2z})=(5.3,\ 1.9,\ 5.3,\ 1.8\ {\rm fm})$. This is qualified to call the $\alpha$ condensation. This configuration is contained in this state by $78$ \%, which is still very large. Furthermore, this container is very close to the one of the Hoyle state, with $(\beta_{1\perp},\beta_{1z})=(5.6,\ 1.4\ {\rm fm})$ in TABLE~\ref{tab:somax_c}. This means that the $0_{V}^+$ state is regarded as the Hoyle analog state, in which the fourth $\alpha$ cluster is also put into the container occupied with the $3\alpha$ clusters in the Hoyle state. The large size of this container indicates that the $4\alpha$ clusters are loosely coupled with each other and configured like a gas. Note that the $4\alpha$ condensate state is also predicted by the $4\alpha$ OCM calculation slightly above the $4\alpha$ threshold, as the $0_6^+$ state. 

These results tell us that the evolution of cluster structures is described by the container evolution with respect to its size and shape. The reason why the container evolution arises is the orthogonality to the lower states, which is explicitly taken into account in the definition of the single configuration ${\widetilde \Phi}_k^{J=0}$ in Eq.~(\ref{eq:so}). The orthogonality condition prevents a higher state configuration from overlapping with the lower-states more compact configurations. It thus plays a role as a repulsive core and is considered to give the container evolution.

In conclusion, I introduced the eTHSR wave function, which comprehensively describes gaslike and non-gaslike cluster states in $\Ox$, standing on the container picture. The evolution of the clustering, as the excitation energy increases, can be obtained by solving the Hill-Wheeler equation, concerning the container parameters. I showed that the evolution of the clustering is caused by the evolution of the container. Not only various $\alpha+\C$ cluster states but also the $4\alpha$ gaslike state are naturally described according to this picture, in which the $\alpha$ clusters are confined into different size and shape containers. In particular, the $4\alpha$ gaslike state is obtained as having the same container for the $4\alpha$ clusters, clearly giving the $4\alpha$ condensate structure. This picture of container evolution is thus a key concept in heavier nuclei, to understand dynamical process of formation of cluster structures, from the ground state to higher excited states.

\vspace{0.2cm}
The author wishes to thank B. Zhou, H. Horiuchi, A. Tohsaki, G. R\"opke, P. Schuck, and T. Yamada for many fruitful discussions. This work is financially supported by the startup fund for the associate professorship of ``Zhuoyue 100'' program, Beihang University.

\end{document}